\documentstyle[aps,twocolumn,epsf]{revtex}
\begin{document}
\draft
\twocolumn[\hsize\textwidth\columnwidth\hsize\csname @twocolumnfalse\endcsname

\title{Melting of a colloidal crystal}
\author{P. S. Kuhn, A. Diehl, Y. Levin, and M. C. Barbosa}
\address{
Instituto de F\'{\i}sica, Universidade Federal
do Rio Grande do Sul\\ Caixa Postal 15051, CEP 91501-970
, Porto Alegre, RS, Brazil}
\maketitle
\begin{abstract}
A melting transition for a system of hard spheres interacting by a repulsive Yukawa potential of DLVO form is studied. To find the location of the phase boundary, we propose a simple theory to calculate the free energies for the coexisting liquid and solid. The free energy for the liquid phase is approximated by a virial expansion. The free energy of the crystalline phase is calculated in the spirit of the Lenard-Jonnes and Devonshire (LJD) theory. The phase boundary is found by equating the pressures and chemical potentials of the coexisting phases. When the approximation leading to the equation of state for the liquid breakes down, the first order transition line is also obtained by applying the Lindemann criterion to the solid phase. Our results are then compared with the Monte Carlo simulations.

\end{abstract}

]

\bigskip
\centerline{\bf 1. Introduction}
\bigskip

The asymmetric polyelectrolyte solutions provide a severe challenge, as well as, a true testing ground for a variety of statistical mechanics theories. Unlike the case of simple symmetric electrolytes \cite{r0}, the phase structure of which is reasonably well understood, it is fair to say that our understanding of charge stabilized colloids is very far from complete. Even such basic questions as what is the form of interactions between strongly charged colloidal particles still remains controversial \cite{r0,r1,Bell,Pa,Vla,Eric}. One thing, however, seems to be well accepted by both the experimentalists and theorists, when the concentration of colloid is sufficiently large it freezes into a crystal \cite{Evert,Ste,Kre,Mei,Ros,Ghosh,Tej,Hone}. The exact mechanism leading to the transition is still not fully understood \cite{r0}. The need to find a resolution to this dilemma is a particularly acute in view of industrial applications of charge stabilized colloids in the design of new materials, such as water soluble paints.

The complete theory of charged colloids would have to include, in addition to polyions, the presence of neutralizing counterions. There have been a number of attempts to study this full system, however, due to severe mathematical complications, no attempt has been made so far to explore the liquid-solid phase transition in highly asymmetric electrolytes \cite{Hone,Lev,r4,Groot}. Instead most approaches rely on using an effective, counterion screened, interaction between the polyions of the DLVO form \cite{Low,Derj}, while completely ignoring the contribution of counterions to the free energy. Although there have been some attempts to compare the phase diagrams derived on the basis of this ``reductionist" approach directly with experiment, we should stress that there is no reason why this type of comparison should at all make sense \cite{Mei}.

At a first order phase transition the pressures and the chemical potentials of the coexisting phases must be equal. Since the biggest contribution to both pressure and chemical potential comes from the entropic motion of the counterions and the interactions between polyions and counterions are strongly coupled, any theory which does not explicitly take into account the presence of counterions is fatally flawed. Nevertheless, in the absence of a complete theory, the reductionist approach has an important role as a testing ground of a variety of theories of the liquid-solid transition. However, the true merit of each theory should be measured by how it compares to the Monte Carlo (MC) simulation of the equivalent reduced model.

Traditionally the theories of liquid-solid transition can be divided into two classes: those based on the density functional methods \cite{Ra} and those based on the modifications of the Lennard-Jones and Devonshire \cite{Le,Ba1} cell theory of liquid. Both methods have their strengths and weaknesses.

In the case of simple electrolytes the cell model has proven to be in a closer agreement with MC \cite{Vega}.
Since the early work of Fuoss, Katchalsky and Lifson \cite{Kat}, the cell theories formed the foundation on which most of the studies of polyelectrolyte solutions were based. The cell models are often indiscriminantly applied to both the solid and liquid phases. In this respect we would like to emphasize that the cell theories entail an underlying periodicity which is valid only in the solid phase. From this perspective our understanding of a crystalline state of colloidal suspensions is quite satisfactory \cite{Low}, and agree with Tejero {\it et al}. approach for point particles \cite{Tej}. What is lacking is an equally good treatment of a disordered state of suspension. Some preliminary approaches in this direction were recently reported in literature \cite{Lev}. In this paper we shall ignore many of the subtleties of the true colloidal suspensions, and confine our attention to a model system of hard spheres interacting by a Yukawa potential of the DLVO form. Furthermore we shall restrict our attention to the strong screening limit, high salt concentration. In this case the repulsive interactions between the equally charged colloids will be sufficiently weak. Thus the virial expansion for, say, the pressure can be truncated at a leading order, allowing us to write an equation of state of a van der Waals form. Unfortunately, as the screening is reduced the truncation of the virial expansion will no longer be valid and an alternative method must be used. The crystalline state of the suspension will be modeled by a modified Lennard-Jones and Devonshire theory.


\bigskip

\centerline{\bf 2. The model}
\bigskip

Our system consist of identical particles of charge $-Zq$ and diameter $\sigma$. The solvent is modelled as an uniform medium of a dielectric constant $D$. In the spirit of reductionist approach the presence of the counterions is neglected beyond their contribution to the screening of the interactions between the colloidal particles. We shall accept that this effective interaction is of a DLVO form \cite{Fis}

\begin{equation}
\label{a1}
W(\vec{r}) = (Zq)_{eff}^2  \; \frac {e^{- \kappa r}} {Dr} \; , \;\;\;\; (Zq)_{eff} \equiv Zq \; \frac {e^{\kappa \sigma /2}} {1+ \kappa \sigma /2} \; ,
\end{equation}

\noindent
where $\kappa$ is an inverse Debye length, which sets the scale for the range of electrostatic interactions. It is convenient for later purposes to write the energy $W$ in the form

\begin{equation}
\label{a2}
W \equiv \epsilon \; \phi(x) \; ,
\end{equation}

\noindent
where

\begin{equation}
\label{b1}
x \equiv r/ \sigma, \;\;\;\; \epsilon \equiv \frac {(Zq)_{eff}^2} {D \sigma} \; ,
\end{equation}

\noindent
and the dimensionless potential $\phi(x)$ is

\begin{equation}
\label{a4}
\phi(x) \equiv \frac {e^{- \kappa \sigma x}} {x} \; .
\end{equation}

\noindent

In the limit $\kappa \sigma \gg 1$ , $\phi(x)$ is extremely short ranged. In this case all the thermodynamic properties of our system will be due to the hard-core repulsion between colloids. On the other hand, if $\kappa \sigma \ll 1$ the thermodynamics will be dominated by the electrostatic interactions.

Our aim will be to find the location of the liquid-solid coexistence region for the two limiting cases. To this end we will need the free energies for both liquid and solid phases. We shall now proceed to construct these free energies.


\bigskip

\centerline{\bf 3.  The fluid phase (\boldmath{$\kappa \sigma \gg 1$})}
\bigskip

The Helmholtz free energy for the fluid phase is constructed as a sum of terms, starting with an ideal gas contribution

\begin{equation}
\label{ff2}
\beta F_f^{id} \equiv N [\log (\rho \Lambda^3) - 1] \; ,
\end{equation}

\noindent
where the thermal de Broglie wavelength is,

\[
\Lambda = \frac {h} {(2 \pi m k T)^{1/2}} \; .
\]

\noindent
The mass of each polyion is denoted by $m$, $h$ is the Planck constant, and $k$ is the Boltzmann constant. The hard-sphere repulsion between the polyions can be included through the Carnahan-Starling contribution to the free energy \cite{Car},

\begin{equation}
\label{b3}
\beta F_f^{CS} \equiv Ny \frac {4-3y} {(1-y)^2} \; , \;\;\;\; y \equiv \frac {\pi \rho^*} {6} \; ,
\end{equation}

\noindent
where $\rho^* \equiv \rho \sigma^3$ is the reduced density.

In the limit $\kappa \sigma \gg 1$ the virial expansion for thermodynamic functions will be composed of rapidly decreasing terms. In this case it will be sufficient to approximate the electrostatic contribution to the free energy by a second virial term,

\begin{equation}
\label{f21}
F_f^e = \frac {N} {2} \int d \vec{r} \; W_{12}(\vec{r}) \; \rho(\vec{r}) \; .
\end{equation}

\noindent
Substituting (\ref{a2}) we find

\begin{equation}
\label{ff3}
F_f^e = \frac {2 \pi \rho^* N} {T^*} \int^{+ \infty}_1 dx \; x^2 \; \phi(x) \; ,
\end{equation}

\noindent
where the reduced temperature is $T^* \equiv 1 / \beta \epsilon = D \sigma / \beta (Zq)_{eff}^2$.
The integral may be easily carried out, leading to 

\begin{equation}
\label{ff5}
\beta F^e_f = \frac {2 \pi \rho^* N} {T^*} \; \frac {e^{- \kappa \sigma} (1 + \kappa \sigma)} {(\kappa \sigma)^2 } \; . 
\end{equation}

\noindent

The full free energy is $F_f = F^{id}_f + F^{CS}_f + F^e_f$. It should be remembered that this expression is only valid for $\kappa \sigma \gg 1$; beyond this limit additional virial terms will contribute to $F^e_f$.



\bigskip
\centerline{\bf 4. The solid phase}
\bigskip

We shall use the following mean  field picture of a colloidal crystal. Each polyion is allowed to move in a cage whose boundaries are defined by its nearest neighbors. However, the particles do not move freely since they interact electrostatically with the neighboring polyions. In the limit $\kappa \sigma \gg 1$ we shall consider the electrostatic interactions with the polyions of the first shell, although in principle other shells can be included reasonably straightforwardly.

The partition function can be written \cite{Hill},

\begin{equation}
\label{c13}
Q= \frac {v_f^N} {\Lambda^{3N}} \; \exp \left(-N \beta U_0/2 \right) \; ,
\end{equation}

\noindent
where $v_f$ is the ``free" or ``effective" volume, available to the center of a polyion inside a cell. Note that no factors of $N!$ are present, since in principle each particle of the lattice can be labeled. The term $N U_0/2$ is the energy of the lattice when all the particles are located at their equilibrium positions.

The free energy of the crystal is

\begin{equation}
\label{s1}
\beta F_s = \frac {N} {2} \beta U_0 - N \ln \left( \frac {v_f} {\sigma^3} \right).
\end{equation}

\noindent
A final approximation of LJD theory is to simplify the geometry of the cell, by taking it to be a sphere with a surface charge $\sigma_e=n_1 (Zq)_{eff}/4 \pi R^2$, where $R \equiv \sigma (\rho_{cp} / \rho)^{1/3}$ is the distance between the nearest neighbors, $\rho_{cp}$ is the ``close-packing" density for a particular crystalline form, and $n_1$ is the number of nearest neighbors. This is refered to as a ``smearing procedure" used elsewhere (as in \cite{McQ}).

The free volume $v_f$ is then simplified to

\begin{equation}
\label{s4}
v_f \equiv 4 \pi \int^{R- \sigma}_0 \exp {[ - \beta (U(r) - U(0))] } r^2 dr \; ,
\end{equation}

\noindent
and $U(r)$ denotes the energy of a particle inside the cell.

In our case we shall assume that the solid state corresponds to an fcc structure with $\rho_{cp}^*=\sqrt{2}$, since this is the transition found in the pure hard-sphere system \cite{Ba1}. 

In order to find the mean field potential exerted on each polyion due to the neighboring polyions we must, in principle, integrate the effective potential (\ref{a1}) over the surface of the sphere. This will require performing some sufficiently messy angular integrals (when the central polyion is displaced from the center of the cell). We can avoid this by observing that the DLVO potential satisfies the Helmholtz equation,

\begin{equation}
\label{s6}
\nabla^2 \phi = \kappa^2 \phi \; .
\end{equation}

\noindent
The required solution is unique, provided we consider the appropriate boundary conditions for the potential $\phi$. These are, continuity at $r=R$, finiteness inside the cell, vanishing value at infinity, and the discontinuity of the normal component of the electric field at $r=R$, due to the presence of a surface charge. We find

\begin{equation}
\label{s7}
\phi_{in}(r) = \frac {1} {4 \pi D} \; n_1 (Zq)_{eff} \; \frac {e^{- \kappa R}} {R} \; \frac {\sinh {(\kappa r)}} {\kappa r} \; , \;\; r<R \; ,
\end{equation}

\noindent                   

\begin{equation}
\label{s8}
\phi_{out}(r) = \frac {1} {4 \pi D} \; n_1 (Zq)_{eff} \; \frac {\sinh {(\kappa R)}} {R} \; \frac {e^{- \kappa r}} {\kappa r} \; , \;\; r>R \; .
\end{equation}

\noindent

The free volume can now be obtained by performing the radial integral (\ref{s4}). We find

\begin{equation}
\label{s13}
\frac {v_f} {\sigma^3} = \frac {4 \pi e^{a_1}} {(\kappa \sigma)^3}  \; I_1 \; ,
\end{equation}

\noindent
where

\begin{equation}
\label{s14}
I_1 \equiv \int^{\kappa R - \kappa \sigma}_0 \exp { \left(- a_1 \frac {\sinh {x}} {x} \right)} \; x^2 \; dx \; , \;\;\;\; x \equiv \kappa r \; ,
\end{equation}

\noindent
and $a_1$ is given by

\begin{equation}
\label{s13a}
a_1 \equiv \frac {1} {4 \pi T^*} \; n_1 \; \kappa \sigma \; \frac {e^{- \kappa R}} {\kappa R} \; .
\end{equation}

\noindent

The above integral is a function of the density and the temperature, and will be performed numerically. Given the free energy, the pressure, $p=- \partial F/ \partial V$, and the chemical potential, $\mu = \partial F/ \partial N$, of the liquid and solid phases can be calculated by a straightforward differentiation. At transition, the pressures and the chemical potentials of the coexisting phases are equal: 

\begin{equation}
\label{sg3}
p_s(\rho_s)=p_f(\rho_f) \; , \;\;\;\; \mu_s(\rho_s)=\mu_f(\rho_f) \; .
\end{equation}

\noindent
Solving these equations numerically the boundary of the coexisting region can be obtained. The coexistence curve $T^*$ {\it versus} $\rho^*$ for three values of $\kappa \sigma$ is ploted on Fig.1. The transition when the only interaction is due to the hard-spheres repulsion is represented by circles. We see that for large values of $\kappa \sigma$ our curve tends to these points, the electrostatic interaction being dominated by the hard sphere contribution. We also observe that in the limit of high temperatures the electrostatic interaction becomes insignificant and the phase transition becomes purely entropic. The assymptotic densities thus obtained ($\rho^*_f = 0.983$, $\rho^*_s = 1.107$) compare quite favorably with those found in MC of a hard sphere mixture ($\rho^*_f = 0.948$, $\rho^*_s = 1.046$) \cite{Hoo}.

\begin{figure}[h]

\vspace*{-4cm}
\begin{center}
\epsfxsize=8.cm
\leavevmode\epsfbox{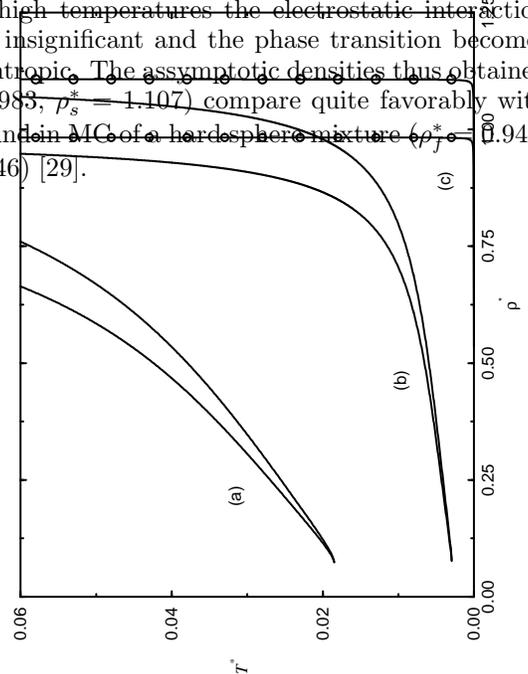}
\end{center}
\vspace*{-0.5cm}
\caption{Fluid-solid coexistence for the repulsive Yukawa system. Three values of $\kappa \sigma$ are shown: $(a) \kappa \sigma = 5$, $(b) \kappa \sigma = 10$, $(c) \kappa \sigma = 100$. The circles represent the hard-sphere transition.}
\end{figure}

The whole curve may be compared with that obtained in the MC simulation of reference \cite{Evert}, and this is done in Fig.2 for $\kappa \sigma = 5$. The agreement is not very good, although the shape of the figure is nearly identical. Once again we would like to remind the reader that our theory remains valid only for sufficiently large values of $\kappa \sigma$. When the screening weakens, the van der Waals approximation will become less reliable, as more and more terms will need to be included in the virial expansion \cite{Lan}.

\begin{figure}[h]

\vspace*{-5cm}
\begin{center}
\epsfxsize=8.cm
\leavevmode\epsfbox{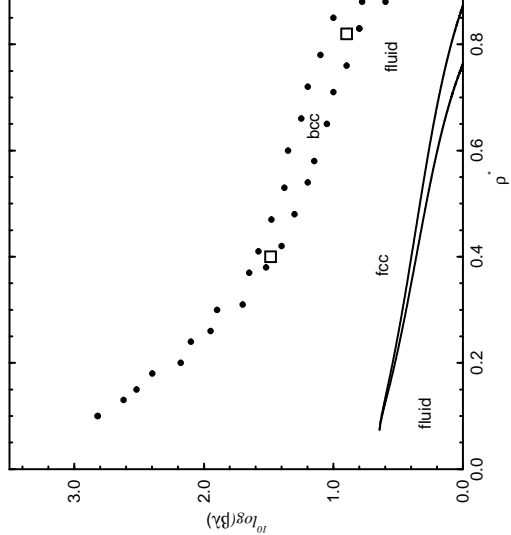}
\end{center}
\vspace*{-0.5cm}
\caption{Fluid-solid coexistence for the repulsive Yukawa system for $\kappa \sigma = 5$. The dotted lines are the MC results of [7]. The squares represent the result for point particles in [10]. We have ploted the results in the units of [7], $\beta \lambda = e^{- \kappa \sigma}/T^*$, to allow comparison.}
\end{figure}

In the absence of a more reliable expression for the free energy for the liquid state, we are, therefore, in need of a drastically different approach. One such method which relies purely on the information obtained from the solid state is the so called Lindemann criterion \cite {Lin} for melting, which states that the crystal will melt when the amplitude of its vibrations become sufficiently large. In order to compare our results with the MC simulations for point particles \cite{Mei} we set $\kappa \sigma = 0$.

The energy of a particle inside the cell is in this case given by

\begin{equation}
\label{p1}
\beta U(r) = {a_1}_p \; \frac {\sinh {(\kappa r)}} {\kappa r} \; ,
\end{equation}

\noindent
where 

\begin{equation}
\label{p2}
{a_1}_p \equiv \frac {1} {4 \pi T^*_p} \; n_1 \; \frac {e^{- \kappa R}} {\kappa R } \; ,  \;\;\;\; T^*_p \equiv \frac {D} {\kappa \beta (Zq)^2} \; .
\end{equation}

\noindent

The semiempirical Lindemann criterion states that the solid will melt when the ratio of the mean-squared displacement($\langle r^2 \rangle$) to the square of the interparticle distance ($a_s^2$) is equal or greater than a constant of order $10^{-2}$. The quantity $a_s \equiv \rho^{-1/3}$ is a geometry independent term, hence applicable to the solid and to a disordered phase as the fluid. This gives an equation for the melting line,

\begin{equation}
\label{p6}
\frac {\langle r^2 \rangle} {a_s^2} = c \; ,
\end{equation}

\noindent
with $c$ a constant with typical values $0.025$, $0.030$, $0.0361$, for Lennard-Jones types of potentials. 

The mean-squared displacement $\langle r^2 \rangle$ is obtained through the canonical average,

\begin{equation}
\label{p7}
\langle r^2 \rangle = \frac {\int^R_0 r^2 \exp {\left(- \beta U(r)\right)} dV} {\int^R_0 \exp {\left(- \beta U(r)\right)} dV} \; .
\end{equation}

\noindent
The integrals must be done numerically, and we write

\begin{equation}
\label{p8}
\langle r^2 \rangle = \frac {I_1} {I_2} \; ,
\end{equation}

\noindent
where

\begin{equation}
\label{p9}
I_1 \equiv \frac {4 \pi} {\kappa^5} \int^{\kappa R}_0 dx \; x^4 \; \exp {\left(-{a_1}_p \frac {\sinh {x}} {x}\right)} \; ,
\end{equation}

\noindent

\begin{equation}
\label{p10}
I_2 \equiv \frac {4 \pi} {\kappa^3} \int^{\kappa R}_0 dx \; x^2 \; \exp {\left(- {a_1}_p \frac {\sinh {x}} {x}\right)} \; ,
\end{equation}

\noindent
and $x \equiv \kappa r$.

Therefore we may draw the melting line for several values of $c$. The resulting curves are the the solid lines shown in Fig.3 for fcc lattice. For typical values of $c$ the curves are located somewhat below the simulation data of reference \cite{Mei} and the theoretical results of Tejero {\it et al}. \cite{Tej}. The agreement is better for higher values of $c$. This can be understood on the following grounds. We have considered only the first shell of neighbors, hence the potential inside a cell is found to be less repulsive than it is in reality. The net effect is that the amplitude of the vibrations of a molecule around its mean location in the cell is overestimated. This increases the value of the ratio $\langle r^2 \rangle /a_s^2$, which is precisely the Lindemann constant. 

This might be clearly seeing if we consider another approximation at this level, which is to solve the integrals above analytically. For this purpose we extend the superior limit of integration to infinity, and substitute the factor $\sinh x/x$ by the first two terms of its series, $1+x^2/6$. Of course this will introduce an error, but we restrict ourselves to the limit of small $\kappa a_s$, namely low salinity. In this case this approximation is acceptable, and the result may be compared with that obtained from the exact numerical calculation. In this limit we find,

\begin{equation}
\label{p12}
\frac {\langle r^2 \rangle} {a_s^2} = \frac {9} {{a_1}_p (\kappa a_s)^2} \; .
\end{equation}

\noindent
From the above expression, the Lindemann criterion, Eq.(21), may be used to draw the melting line. The results are ploted as the dashed lines in Fig.3 for fcc lattice. We see that for small values of $\kappa a_s$ the curves are close to the exact ones. Even though this simplification gives a worst agreement with the simulation, it has the advantage of allowing us to write an analytical expression for the melting temperature as a function of the density. From the equation of melting, (\ref{p6}), and the expression for the relative mean squared displacement, Eq.(\ref{p12}), we obtain the melting temperature, 

\begin{equation}
\label{pp1}
{T^*_p} (\kappa a_s) = \frac {c \, n_1} {36 \pi} \; \frac {e^{- \kappa R}} {\kappa R} \; (\kappa a_s)^2 \; .
\end{equation}

\noindent

From the above equation we can see that the transition temperature depends on the coefficient $c n_1$. Therefore an increase in the temperature would be obtained both by increasing the Lindemann constant or by considering further neighbors and replacing $n_1$ by $n_{eff}>n_1$.

\begin{figure}[h]

\vspace*{-4cm}
\begin{center}
\epsfxsize=8.cm
\leavevmode\epsfbox{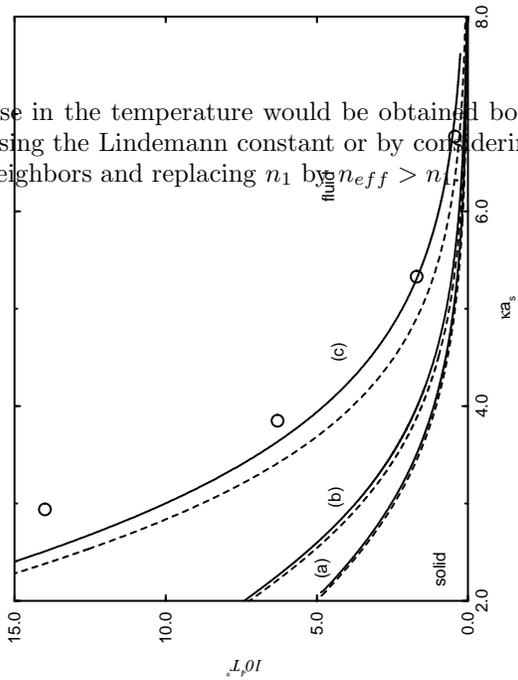}
\end{center}
\vspace*{-0.5cm}
\caption{Melting lines for the repulsive Yukawa system as obtained from Lindemann criterion. We have point particles and the fcc symmetry. Solid lines are the result of exact calculation, and dashed lines are obtained from the gaussian approximation (see text). The full circles are the simulation coexistence points of [10]. The values of the Lindemann constant were taken as $0.025$ in line $(a)$, $0.0361$ in line $(b)$, and $0.09$ in line $(c)$.}
\end{figure}

The Lindemann criterion may also be used for particles of finite diameter, and we do this for $\kappa \sigma = 5$, for comparison with the MC results of \cite{Evert}. In this case we have

\begin{equation}
\label{p13}
\langle r^2 \rangle = \frac {I_1} {I_2} \; ,
\end{equation}

\noindent
where

\begin{equation}
\label{p14}
I_1 \equiv \frac {4 \pi} {\kappa^5} \int^{\kappa R - \kappa \sigma}_0 \; dx \; x^4 \; \exp \left(- a_1 \frac {\sinh x} {x} \right) \; ,
\end{equation}

\noindent

\begin{equation}
\label{p15}
I_2 \equiv \frac {4 \pi} {\kappa^3} \int^{\kappa R - \kappa \sigma}_0 \; dx \; x^2 \; \exp \left(- a_1 \frac {\sinh x} {x} \right) \; ,
\end{equation}

\noindent
and $a_1$ is defined in Eq.(\ref{s13a}).

With the above results the melting line is constructed, and the results are shown as the solid lines in Fig.4. The dotted line is the curve of \cite{Evert}. We now consider a simplified version as was done in the point particle case, namely to evaluate the integrals in the expression for $\langle r^2 \rangle$ from $0$ to $\infty$, and to replace $\sinh x/x$ by $1+x^2/6$. The same range of validity is understood. We obtain as before

\begin{equation}
\label{p18}
\kappa^2 \langle r^2 \rangle = \frac {9} {a_1},
\end{equation}

\noindent
but remember that $a_1$ is now given by Eq.(\ref{s13a}). We may draw the corresponding melting lines for several values of the constant $c$, and the results are the dashed lines in Fig.4. With the above expression for $\kappa^2 \langle r^2 \rangle$, the equation for the melting, (\ref{p6}), we obtain the melting temperature as for point particles, $T^* = {T^*_p} \kappa \sigma$. Note the difference in the definitions of reduced temperatures for finite size and for point particles.

We note that once again the best agreement with computer experiment is found for $c=0.09$. Furthermore the approximated form for the melting temperature produces a better fit to MC, sugesting that the cell model overestimates the effects of confinement.

\begin{figure}[h]

\vspace*{-4cm}
\begin{center}
\epsfxsize=8.cm
\leavevmode\epsfbox{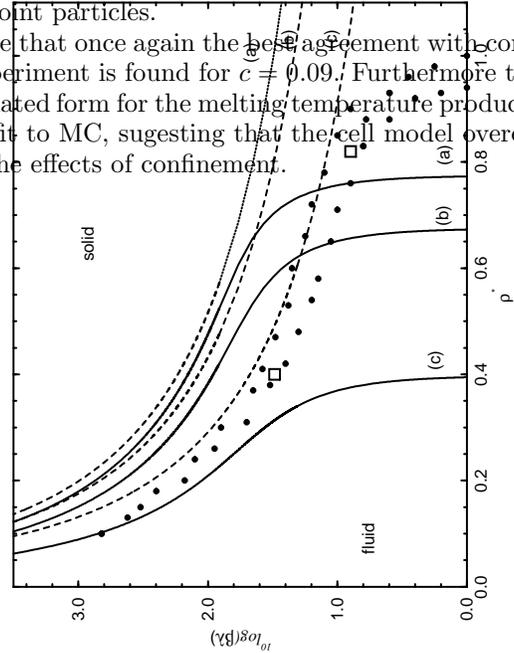}
\end{center}
\vspace*{-0.5cm}
\caption{Melting lines for the repulsive Yukawa system as obtained from Lindemann criterion. The particles have finite size($\kappa \sigma = 5$) and the symmetry is fcc. Solid lines are the result of exact calculation, and dashed lines are obtained from the gaussian approximation (see text). We have adopted the units of [7] to allow comparison. The dotted lines are the MC results of [7]. The squares represent the result for point particles in [10]. The values of the Lindemann constant were taken as $0.025$ in line $(a)$, $0.0361$ in line $(b)$, and $0.09$ in line $(c)$.}
\end{figure}



\bigskip
\centerline{\bf 5. Conclusions}
\bigskip

We have studied a melting transition in a system composed of hard spheres interacting by a repulsive Yukawa potential. A simple theory to calculate the free energies of the coexisting liquid and solid phases was proposed. Since our equation of state for the liquid is based on the virial expansion, it is expected to work well only in the limit of strong screening, since in this limit it is sufficient to keep the first virial term. However, if a more reliable equation for the liquid state can be found it should, in principle, be possible to extend our results over the full range of salinity. For the solid phase, we introduced a cell model based on the Lennard-Jones and Devonshire theory. From the free energy we derived expressions for the pressure and chemical potential.

The coexistence curve was obtained by equating the pressures and the chemical potentials of both phases. We compared our results with simulations in Fig.2. Eventhough the results are qualitatively similar, the actual value of the densities for the coexisting phases are quite different. This is most likely due to the breakedown in our approximation leading to the equation of state for the liquid.

In the absence of a better liquid state theory, we have relied on the Lindemann criterion to estimate the location of the melting line for low salinity. With a suitable choice for the Lindemann parameter, a reasonable fit to the MC data was found. Unfortunately this parameter is significantly larger than is usual for the systems with short range interactions.

\bigskip
\centerline{\it ACKNOWLEDGMENTS}
\bigskip

This work was supported in part by CNPq - Conselho Nacional de
Desenvolvimento Cient\'{\i}fico e Tecnol\'ogico and FINEP -
Financiadora de Estudos e Projetos, Brazil.

\end{document}